\documentclass[prb,aps,12pt,a4paper,showpacs]{revtex4}

\usepackage{epsfig}

\begin{document}

\title{ 
        Density of states for almost diagonal random matrices
      }

\author{Oleg Yevtushenko}
\email{bom@ictp.trieste.it}
\affiliation{The Abdus Salam ICTP, Strada Costiera 11, 34100, Trieste, Italy}

\author{Vladimir E. Kravtsov}
\email{kravtsov@ictp.trieste.it}
\affiliation{ The Abdus Salam ICTP, Strada Costiera 11, 34100, Trieste, Italy \\
              Landau Institute for Theoretical Physics, 2 Kosygina st.,
              117940 Moscow, Russia
            }

\date{\today}

\begin{abstract}
We study the density of states (DOS) for disordered systems whose spectral 
statistics can be described by a Gaussian ensemble of almost diagonal 
Hermitian random matrices. The matrices have independent random 
entries $ \, H_{i \geq j} \, $ with small off-diagonal elements: $ \, 
\langle|H_{i \neq j}|^{2} \rangle \ll \langle|H_{ii}|^{2} \rangle \sim 
1 $. Using the recently suggested method of a {\it virial expansion in 
the number of interacting energy levels} (Journ.Phys.A {\bf 36}, 8265), 
we calculate the leading correction to the Poissonian DOS in the cases of 
the Gaussian Orthogonal and Unitary Ensembles. We apply the general formula 
to the critical power-law banded random matrices and the unitary 
Moshe-Neuberger-Shapiro model and compare DOS of these models.
\end{abstract}

{}

\maketitle

Recently, extensive attention has been devoted to unconventional random 
matrix theories (RMTs) that interpolate between the Wigner-Dyson RMT and 
banded RM (BRM) with the (almost) Poissonian level statistics and can be 
used as a helpful tool to explore the localization transition. One of 
these models is the {\it power law banded random matrix} (PLBRM) theory 
\cite{MF,KM,ME} for which the variance of the off-diagonal elements reads
\begin{equation}\label{VarPL}
  {\rm PLBRM:} \quad
  \langle | V_{ij}|^{2} \rangle = { 1 \over 2 } \frac{1}{ 1 + 
                              \left( 
                        \frac{N}{\pi} \sin \left( \frac{\pi}{N} |i-j| \right) 
                              \right)^{2\alpha} / \, b^{2\alpha}
                                          } \, .
\end{equation}
It is nearly constant inside the band $ \, |i-j| < \lambda \sim b $, and decreases
as a power-law function $\langle |V_{ij}|^{2}\rangle \sim 1/|i- j|^{-2 \alpha}$
for $ \, |i-j|> \lambda $.  Eq.(\ref{VarPL}) is written for periodic boundary
conditions of the PLBRM Hamiltonian. The special case $ \, \alpha=1 \, $ is
relevant for description of critical systems with multifractal eigenstates
\cite{MF,KM,ME,AltLev,Levitov}, in particular for systems at the Anderson
localization-delocalization transition point. On the other hand, it has been
conjectured \cite{KrTs} that the spectral statistics of critical PLBRM with large
$ \, b \, $ can be mapped onto the Calogero-Sutherland model (CS)
[\onlinecite{CS}] at low temperature where instead of the spectral problem one
studies the statistics of interacting (for the real off-diagonal elements in PLBR)
or non-interacting (for the complex off-diagonal elements in PLBR)  fermions in a
parabolic confinement potential.  The case $\alpha >1$ corresponds to the power-law
localization which can be found in certain periodically driven quantum-mechanical
systems \cite{KickRotPLRM}. If $ \, \alpha \le 1/2 \, $ the spectral statistics of
PLBRM approaches the Wigner-Dyson universality class with $ \, \beta = 1 \mbox{ or
} 2 $.

The exactly solvable model of Moshe, Neuberger and Shapiro (MNS) also 
incorporates both the Poissonian and the Wigner-Dyson level statistics \cite{MNS}.
The probability distribution of the Hamiltonian $ \, \hat {\cal H} \, $ in MNS is 
given by $P(\hat{{\cal H}})=\int {\rm d}\hat{U} \ {\cal P}_{\hat U} ( \hat {\cal
H} ) $, where
\begin{equation}\label{MNS-def}
 {\cal P}_{\hat U} ( \hat {\cal H} ) \propto \exp \Biggl( - {\rm Tr} 
            \hat{{\cal H}}^2 - 
            \left( \frac{N}{ 2 \pi b } \right)^2 {\rm Tr} \Bigl( [ \hat{U}, 
            \hat{{\cal H}} ] [ \hat{U}, \hat{{\cal H}} ]^{\dagger} \Bigr) 
            \Biggr) \, ; 
\end{equation}
the matrix $ \, \hat U \, $ is either unitary for complex Hermitian matrices $ \, 
\hat {\cal H} \, $ (the unitary MNS) or orthogonal for real symmetric matrices $ \, 
\hat {\cal H} \,$ (orthogonal MNS), and $ \, {\rm d}\hat{U} \, $ is the Haar measure. 

The spectral properties of the unitary MNS turn out to be equivalent to a 
system of noninteracting one dimensional ($ 1d $) fermions
in a parabolic confinement\cite{MNS}.  The spectral statistics of the orthogonal
MNS coincides \cite{Garcia} with the statistics of $ \, 1d \, $ fermions in a parabolic
potential with the long-range attractive interaction $\propto (x_{i}-x_{j})^{-2}$.  
This model of strongly correlated fermions is a particular case of the
Calogero-Sutherland model which has been intensively studied as a toy model for 
the fractional statistics. In both cases the parameter $ \, b \, $ of MNS 
corresponds to the inverse temperature of CS: $ \, b \sim 1/T_{CS} $.

The connection between the two models is especially clear in the unitary 
case where the unitary matrix $\hat U=M\, diag\{e^{i\varphi_{i}} 
\}\,M^{\dagger}$ can be diagonalized by a unitary transformation. Then 
the variances of $ \, V_{i,j} = \left( M^{\dagger} \hat{{\cal H}} M 
\right)_{i,j} \, $ in MNS are given by:
\begin{equation}\label{VarMNS}
  {\rm MNS:} \quad
  \langle | V_{ij}|^{2} \rangle = { 1 \over 2 } \frac{1}{ 1 + 
                              \left( 
                        \frac{N}{\pi b}\right)^{2}\, \sin^{2} \left( 
\frac{\varphi_i-\varphi_j}{2}\right)
                                          } \, .
\end{equation}
One can easily see that Eq.(\ref{VarMNS}) coincides with Eq.(\ref{VarPL}) 
at $\alpha=1$ if the phases $\varphi_{n}=2\pi n/N$ are arranged as an 
ordered array on a circle. In general the MNS model can be considered as 
an extension of the PLBRM model for the case of the arbitrary arrangement 
of phases $\varphi_{n}$ homogeneously distributed over the circle.

The following formula is valid  to calculate the averaged value of an 
observable $ \, A(\hat{H}) \, $ which is invariant under the transformation
$ \, \hat{H} \to M^{\dagger} \hat{{\cal H}} M $:
\begin{equation}\label{MNSaver}
   \langle \langle A \rangle_{\hat{H}} \rangle_{\hat{U}} \equiv
       { 
         \int \langle A \rangle_{\hat{H}} \, P(\{\varphi_{i} \})\, {\rm D} 
\{ 
\varphi_i \} \over 
         \int P(\{\varphi_{i}\}) \, {\rm D} \{ \varphi_i \}
       } \, .
\end{equation} 
Here $ \, P(\{\varphi \}) \, $ is the joint probability distribution of  
phases [\onlinecite{MNS}]:
\begin{equation} \label{JPD}
   P(\{\varphi\}) \sim \prod_{i>j}
                    \frac{ \sin^2 \left( \frac{\varphi_i-\varphi_j}{2} 
                       \right) }{ 1 +
                              \left(
                        \frac{N}{\pi b}\right)^{2}\, \sin^{2} \left(
   \frac{\varphi_i-\varphi_j}{2}\right)} \, ,
\end{equation}
and $\langle A \rangle_{\hat{H}}$ stands for the averaging over the 
Gaussian random matrix $\hat{H}$ with entries having zero mean value and 
the variance given by Eq.(\ref{VarMNS}).

The two-point correlation function, which follows from Eq.(\ref{JPD}) after the
integration over all but two phases, was calculated by Gaudin with the help of the
model of free non-interacting fermions with a linear spectrum \cite{Gaudin}:
\begin{equation} \label{PhR2}
   {\cal R}_2 ( s ) = 1 - \frac{1}{( 2 \pi b )^2 } 
                   \left|
                \int_{ - \log \left( e^{2 \pi b} - 1 \right) }^{\infty}
                   \frac{ e^{ {\bf i} { \omega s \over b } } \, {\rm d} \omega }
                        { e^{\omega} + 1 }
                   \right|^2 \, , \quad
  s=\phi_i-\phi_j\equiv (\varphi_i-\varphi_j)(N/2\pi) \, .
\end{equation}
If $ \, | s | \gg b $, the correlation function is almost constant $ \, {\cal R}_2 
\Bigl( | s | \gg b \Bigr)  \to 1 $. There is a repulsion between phases at a small 
scale controlled by $ \, b $: $ \, {\cal R}_2 \Bigl( | s | \ll b \Bigr) \sim ( s / 
b )^2 $.

For $ \, b \gg 1 \, $ the spectral statistics of the critical PLBRM with $ \, \alpha 
= 1 \, $ and MNS are asymptotically the same and at $b\rightarrow \infty$ they approach
the Wigner-Dyson statistics \cite{KM,ME,KrTs}. This is because the phase repulsion
in MNS is strong at large $ \, b $.  The phases $ \, \phi_{i,j} \, $ form an
approximately equidistant lattice-like structure \cite{KM}.  In the opposite case
$ \, b \ll 1 \, $, the phase repulsion in MNS is weak and the phases $ \,
\phi_{i,j} \, $ do not form a regular structure. Disorder in the phase
arrangements at a small distances $ \, | \phi_i - \phi_j | \sim 1 \, $ may become
especially important. Therefore, there is no evident correspondence between
critical PLBRM and MNS at $b\ll 1$ even though both ensembles have the same first
correction to the Poissonian level rigidity \cite{KrMac,ME} and the numerics have
revealed a relatively small difference in the level rigidity of PLBRM and MNS at $
\, b \sim 1 $ [\onlinecite{KrMac}].

The progress in BRM and PLBRM theories became possible because of mapping
\cite{MF91,MF} onto the nonlinear supersymmetric sigma-model \cite{Efetov} that
allowed to obtain rigorous results by using various powerful methods of the field
theory. However, the sigma-model always starts from delocalized (i.e. diffusive)
modes and such mapping is only justified if the bandwidth $ \, \lambda \gg 1 $. In
the opposite case where all the off-diagonal matrix elements are parametrically
small compared to the diagonal ones and the system is close to the localization,
no field-theoretical approach is known so far. Yet such {\it almost diagonal} RMT
may possess nontrivial properties because of the slow decay of the off-diagonal
matrix elements $ \, \langle |H_{ij}|^{2} \rangle \, $ with increasing $ \, |i-j|
$. For instance it is of fundamental interest to study the spectral statistics in
systems with {\it power-law localization} that takes place in the power-law banded
random matrix ensembles at $ \, \alpha > 1 $. Another problem to study is the
critical almost diagonal PLBRM. It is known that the eigenvectors of PLBRM with $
\, \alpha = 1 \, $ remain multifractal for an arbitrary small value of $ \,
\lambda \, $ [\onlinecite{Levitov}]. This means that the typical eigenfunction is
{\it extended} though very sparse at small $ \, \lambda $. Their fractal
dimensions are small as compared to the dimension $ \, d = 1 \, $ of the
underlying chain with the long range hopping. Thus almost diagonal PLBRMs may
display the {\it localization-delocalization transition} with changing the
exponent $ \, \alpha \, $ as well as their large bandwidth counterpart \cite{MF}.

Recently, we have suggested a new method that allows to study spectral statistics
of a disordered system described by an ensemble of the almost diagonal random
matrices \cite{Machin}. It is a {\it virial expansion in the number of interacting
energy levels}. Unlike the field-theoretical approach, the virial expansion starts
from the Poissonian statistics and yields a regular expansion in powers of the
small parameter controlling the ratio of the off-diagonal elements to the diagonal
ones $ \, \langle | H_{i \ne j} |^2 \rangle / \langle H_{ii}^2 \rangle \sim b^2
\ll 1 $. The expansion has been represented by the summation of diagrams which are
generated with the help of the Trotter formula. A rigorous selection rule has been
established for the diagrams, which allows to account for exact contributions of a
given number of resonant and non-resonant interacting levels. The method offers a
controllable way to find an answer to the question when a weak interaction of
levels can drive the system from localization toward criticality and
delocalization. An example of the spectral form-factor has been considered for a
generic dependence of the variance $ \, \langle | H_{i \ne j} |^2 \rangle \, $ on
the difference $ \, i - j $. It has been shown that a term of the order of $ \,
b^{c-1} \, $ is governed by the interaction of $ \, c \, $ energy levels. The
general theory has been applied to the Rosenzweig-Porter \cite{RosPort} model and
to the critical PLBRM.

In the present paper, we continue studying the spectral statistics with the help
of the virial expansion. We calculate the density of states (DOS) for the
ensembles of the Gaussian almost diagonal random matrices. Based on the detailed
presentation of the method in the paper [\onlinecite{Machin}], we will explain the
corresponding diagrammatic technique for DOS.  We derive the leading correction to
the Poissonian DOS for the models of critical PLBRM and MNS.

Let us consider a Hermitian random matrix (RM) of size $ \, N \times N $, $ \, N
\gg 1 $, from a Gaussian ensemble. We assume that the matrix entries are random
and independent. The RM is the Hamiltonian $ \, \hat H \, $ of the matrix
Schr{\"o}dinger equation $ \, \hat H \psi_n = \epsilon_n \psi_n $, where $ \,
\epsilon_n \mbox{ and } \psi_n \, $ are the eigenvalues and eigenvectors,
respectively. We define statistical properties of the matrix entries as:
\begin{equation}\label{model}
   \langle H_{i,j}   \rangle = 0 \, ; \quad
   \langle H_{i,i}^2 \rangle = { 1 \over \beta } \, ; \qquad
   \langle | H_{i,j} |^2 \rangle = b^2 \, {\cal F}( | i - j | ) \, , \ i \ne j \, ;
\end{equation}
where $ \, {\cal F}(|i-j|) > 0 \, $ is a smooth function of its argument, and the
parameter $ \, b \, $ is small:
\[
   b \ll 1 \, .
\]
The condition $ \, b \ll 1 \, $ means that RM is {\it almost diagonal}. The
parameter $ \, \beta \, $ corresponds to the Dyson symmetry classes: $ \, \beta_{GOE} 
= 1 $, $ \ \beta_{GUE} = 2 $. The brackets $ \, \langle \ldots \rangle \, $ denote the 
ensemble averaging. 

We will study the spectral properties of the system concentrating on the ensemble
averaged density of states:
\begin{equation}\label{DOS-E}
  \rho(E) = \langle \sum_n \delta( E - \epsilon_n ) \rangle \, .
\end{equation}
For almost diagonal RM the representation of spectral statistics in the time
domain is more convenient \cite{Machin} therefore we explore below the Fourier
transformed DOS as a function of time:
\begin{equation}\label{DOS-t}
  C(t) = \langle \, {\rm Tr} \, e^{ \, {\bf i} \hat{H} t} \, \rangle \, .
\end{equation}

We start with a brief explanation of {\it the method of the virial expansion} that
has been developed in detail in Ref.[\onlinecite{Machin}]. As far as we
investigate the properties of almost diagonal RMs the Hamiltonian can be naturally
divided into a diagonal part $ \, \hat{H}_\varepsilon \, $ and a matrix of hopping
elements $ \, \hat{V} $:
\begin{equation}\label{sep}
  \hat{H} \equiv \hat{H}_\varepsilon + \hat{V} \, .
\end{equation}
For a strictly diagonal matrix ($\hat{V} = 0$) the Poissonian DOS $ \, C_P(t) \, $
is calculated straightforwardly from Eq.(\ref{DOS-t}):
\begin{equation}\label{Cp}
  C_P (t) = N e^{ - \frac{t^2}{2\beta} } \, .
\end{equation}

It follows from the definition (\ref{model}) that the hopping elements $ \,
H_{i,j} \equiv V_{i,j} \sim b \, $ are small compared to the diagonal ones $ \,
H_{i,i} \equiv \varepsilon_i \sim 1 $. However, a direct expansion of the
exponentials $ \, e^{ {\bf i} \, ( \hat{H}_\varepsilon + \hat{V} ) t } \, $ in
Eq.(\ref{DOS-t}) in terms of $ \, \hat{V} \, $ would involve serious difficulties
because the matrices $ \, \hat{H}_\varepsilon \mbox{ and } \hat{V} \, $ do not
commute with each other. One possible way to overcome these problems is to
represent $ \, e^{ {\bf i} \, ( \hat{H}_\varepsilon + \hat{V} ) t } \, $ as a
product of exponentials containing matrices $ \, \hat{H}_\varepsilon \mbox{ and }
\hat{V} \, $ separately. We do this using the {\it the Trotter formula} \cite{Trt}:
\begin{equation}\label{Trott}
  e^{\hat{A}+\hat{B}} = \lim_{n\to\infty} \left( e^{\hat{A}/n} \, e^{\hat{B}/n} 
                                          \right)^n
       \quad \Rightarrow \quad
  e^{ \, {\bf i} \, \hat{H} t } =
     \lim_{n\to\infty}\prod_{p=1}^{n}
               \left(  e^{ \, {\bf i} \, \hat{H}_\varepsilon t/n }
                       e^{ \, {\bf i} \, \hat{V}             t/n }
               \right) \, .
\end{equation}
In order to obtain corrections to $ C(t)$ proportional to $ (bt)^{m}$ one has to
expand $ \, m \, $ different exponentials in the infinite product in the r.h.s. of
Eq.(\ref{Trott}), $ \, e^{ \, {\bf i} t \hat{V}/n} \simeq 1 + {\bf i} \frac{t
\hat{V}}{n} $, setting in the rest $ \, n - m \, $ exponentials $ \, \hat{V} \to 0
$, and to perform the Gaussian averaging over $\hat{H}_{\varepsilon}$ and
$\hat{V}$.  The terms with an odd power $ \, m \, $ give zero after the disorder
averaging. Thus, the power $ \, m \, $ must be even and we can substitute $ \, m
\to 2 k $.

There are $ \, \frac{n!}{2k!(2k-n)!}\, $ ways to choose $2k$ exponentials to be
expanded in $ \, \hat{V} \, $ from the r.h.s. of Eq.(\ref{Trott}). Therefore,
before taking the limit $ \, n \to \infty $, one has to account for all possible
different arrangements of the expanded exponentials which results in a {\it
summation over the Trotter variables} $ \, p_1, p_2, \ldots p_{2k-1} $; $ \,
\sum_{l=1}^{2k-1} p_l \equiv n $. Each discrete variable $ \, p_l \, $ denotes the
number of {\it successive} exponentials $ \, e^{ \, {\bf i} \, \hat{H}_\varepsilon
t/n } \, $ fused together:
\begin{eqnarray}
  & \langle
   \ldots e^{ \, {\bf i} \, \hat{H}_\varepsilon t/n } \left( 
                                                     1 + {\bf i} \frac{t \hat{V}}{n} 
                                                      \right)
      \underbrace{
          \times \
          e^{ \, {\bf i} \, \hat{H}_\varepsilon t/n } \ldots 
          e^{ \, {\bf i} \, \hat{H}_\varepsilon t/n }
          \ \times 
                 }_{ p_l \mbox{ exponentials where } \hat{V} \to 0 }        
                                                     \left( 
                                                     1 + {\bf i} \frac{t \hat{V}}{n} 
                                                      \right)
   e^{ \, {\bf i} \, \hat{H}_\varepsilon t/n } \ldots
  \rangle = \cr \cr
  & = \langle
   \ldots e^{ \, {\bf i} \, \hat{H}_\varepsilon t p_{l-1}/n } \left( 
                                                      1 + {\bf i} \frac{t \hat{V}}{n} 
                                                              \right)
          e^{ \, {\bf i} \, \hat{H}_\varepsilon t p_{l}/n }   \left( 
                                                      1 + {\bf i} \frac{t \hat{V}}{n} 
                                                              \right)
   e^{ \, {\bf i} \, \hat{H}_\varepsilon p_{l+1}t/n } \ldots
  \rangle  \, . 
  \nonumber
\end{eqnarray}
We can introduce scaled variables $ \, Y_l = p_l / n \, $ converting the summation
over $ \, p_l \, $ to the integration over $ \, Y_l \, $ and eliminating the
parameter $ \, n \, $ from further calculations. The resulting expression must be
averaged over the diagonal elements $ \, \varepsilon_i \, $ and yields the
integral $ \, {\cal I}_{\beta} (t,k) \, $ which depends on the time $ \, t $, on
the power $ \, k \, $ and on the parameter $ \, \beta $.

For almost diagonal RMs, the higher number of the interacting energy levels the
smaller is the correction to the Poissonian level statistics governed by that
interaction. In particular, to find {\it the leading in $ \, b \, $ correction $
\, C_1(t) \, $ to $ \, C_P (t) $},
\[
  C(t) \simeq C_P (t) + C_1(t) + \ldots  \, ; \quad 
  C_1( t, \, b \ll 1 ) \ll C_P (t) \, ,
\]
we retain in the obtained series all terms that correspond to an interaction of
two different species of the diagonal elements $ \, \varepsilon_{i} \, $ and $ \,
\varepsilon_{j} \, $ via the hopping elements $ \, V_{i,j} \, $ and $ \, V_{j,i}
\, $ for any indices $ \, i \ne j \, $ in the range from 1 to $ N $. Then at fixed
$ \, k \, $ we find the following contribution to the correction $ \, C_1(t) $:
\[
  C_1^{(k)} =
  ( \, {\bf i} \, t \, )^{2k} \, {\cal I}_{\beta}(t,k) \, 
          \sum_{i \ne j}^N \langle \Bigl( V_{i,j} V_{j,i} \Bigr)^k \rangle \, .
\]
After integration over $ \, 2k - 2 \, $ Trotter variables $\, \{ Y_1, Y_2, \ldots
Y_{2k-2} \} $, $ \, {\cal I}_{\beta} \, $ can be simplified to one-dimensional integral:
\begin{equation}\label{int}
   {\cal I}_{\beta}(t,k) = \, \frac{e^{-\frac{t^2}{4\beta}}}{k! \, (k-1)!} \,
      \int_0^{1/2} (1/4 - Y^2)^{k-1} \, e^{-(tY)^2/\beta} \, {\rm d} \, Y \, .
\end{equation}
In accordance with the definition (\ref{model}), the Gaussian average of $ \, \langle 
\Bigl( V_{i,j} V_{j,i} \Bigr)^k \rangle \, $ can be transformed to the following form:
\begin{equation}
   \langle \Bigl( V_{i,j} V_{j,i} \Bigr)^k \rangle = 
                            \, {\cal K}_{\beta}(k) \times  b^{2k} \, {\cal F}^k (|i-j|) \, ,
\end{equation}
where $ \, {\cal K}_{\beta}(k) \, $ is the combinatorial factor. Due to the 
Wick theorem, it is equal to factorials:
\begin{equation}\label{Comb}
   {\cal K}_{\beta}(k) = \left\{
    \begin{array}{l}
       (2k-1)!! \, , \ \beta = 1 \, ; \cr
       k!       \, , \ \beta = 2 \, .
    \end{array}
                         \right.
\end{equation}
We have to sum the contributions $ \, C_1^{(k)} \, $ over the parameter $ \, k \,
$ at the end, $ \, C_1 = \sum_k C_1^{(k)} $. This summation yields the answer for
$ \, C_1(t) \, $ as a series in powers of the product $ \ ( b \, t ) $:
\begin{figure}
\unitlength1cm
\begin{picture}(13,3.0)
   \epsfig{file=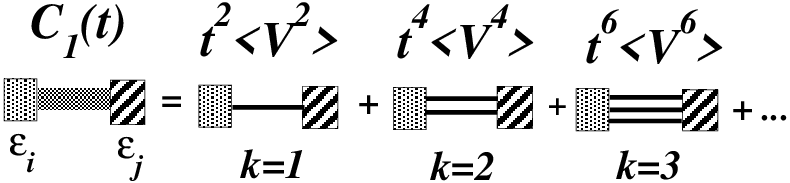,angle=0,width=13cm}
\end{picture}
\vspace{0.5cm}
\caption{
\label{2ColSer}
Graphic illustration of the series (\ref{DeltaC}) for the leading correction $ \,
C_1(t) \, $ to the Poissonian DOS $ \, C_P(t) $. Shadowed boxes mark the energy
levels with different shadowing (``colors'') for $ \, \varepsilon_i \, $ and $ \,
\varepsilon_j $. In each diagram with a given $ \, k $, they are connected by the
$k$ interaction lines which are associated with the factor $ \, t^{2k} \langle
\hat{V}^{2k} \rangle \sim ( t \, b )^{2k} $.
}
\end{figure}
\begin{equation}\label{DeltaC}
  C_1(t) = N\sum_{k=1}^{\infty} (-1)^k \ ( \, b \, t \, )^{2k} \ 
                                 {\cal I}_{\beta}(t,k) \ {\cal K}_{\beta}(k) \ 
                                 {\cal R}_N(k) \, ; \quad
   {\cal R}_N(k) \equiv \frac{1}{N}\sum_{i \ne j}^N {\cal F}^k (|i-j|) \, ,
\end{equation}
\noindent
see a graphic presentation in Fig.\ref{2ColSer}. We will show below that $ \, C_1
\, $ is not larger than $ \, O(b^1) $. We emphasize that neither the combinatorial
factor $ \, {\cal K}_{\beta}(k) \, $ nor the integral over the Trotter variables $
\, {\cal I}_{\beta}(t,k) \, $ depend on the correlation function $ \, {\cal F} ( |
i - j | ) $. Thus, they are universal. The factor $ \, {\cal R}_N (k) \, $ is, on
the contrary, model dependent. It arises because of summation of the product of
the correlation functions $ \, {\cal F} (|i-j|)  \, $ over the indices $ \, i \, $
and $ \, j \, $ and is not universal. If we associate the Hamiltonian $ \, \hat H
\, $ with a one-dimensional chain having a long range hopping between different
sites the summation over $ \, i \, $ and $ \, j \, $ turns out to be the summation
in the real space along the sites of the chain. As the function $ \, {\cal F}
(|i-j|) \, $ depends only on the difference of $ \, | i - j | \, $ the leading part 
of the real space sum is:
\begin{equation}\label{RSS} 
   R_N(k) \simeq 2 \sum_{m=1}^{\infty} {\cal F}^k(m) + O(1/N) \, .
\end{equation}
In what follows we will neglect the correction of the order $ \, O(1/N) \, $ to $ 
\, R_N(k) $.

To simplify further analysis of DOS we insert Eqs.(\ref{int},\ref{Comb},\ref{RSS})
into the series (\ref{DeltaC}) and change the summation order. At first, we sum
over the power $ \, k \, $ getting the answer for $ \, C_1(t) \, $ as a
one-dimensional series over $ \, m $, i.e., as the sum in the real space:
\begin{equation}\label{DeltaCsimp}
  C_1(t) = -2 N e^{ -\frac{t^2}{4\beta} } \sum_{m=1}^{\infty} Z( b \, t, m) \, 
                 \tilde{{\cal I}}_{\beta} \Bigl( t, Z( b \, t, m) \Bigr) \, , \quad
  Z( b \, t, m) \equiv ( b \, t )^2 {\cal F}(m) \, ;
\end{equation}
\begin{equation}\label{IntSimp}
  \tilde{{\cal I}}_{\beta} \Bigl( t, Z \Bigr) \equiv 
    \int_0^{1/2} \!\! {\rm d} \, Y \ e^{ - \Bigl[ \frac{(tY)^2}{\beta} + 
                                                           (\frac{1}{4}-Y^2)Z \Bigr] } 
               \, \times \, 
     \left\{
       \begin{array}{l}
         I_0 \Bigl( (\frac{1}{4}-Y^2)Z \Bigr) - I_1 
                  \Bigl( (\frac{1}{4}-Y^2)Z \Bigr) , \ \beta = 1 \, ; \cr
         1, \ \beta = 2 \, .
       \end{array}
     \right.
\end{equation}
An analytical integration over $ \, Y \, $ is easily doable at $ \, \beta = 2 \, $
and the integral (\ref{IntSimp}) simplifies to the error function:
\begin{equation}\label{IntSimpGUE}
  \tilde{{\cal I}}_{\beta=2} \Bigl( t, Z \Bigr) = e^{-Z/4} 
       \sqrt{ \frac{\pi}{2(t^2-2Z)} } \ {\rm erf} 
             \left( \sqrt{ \frac{t^2-2Z}{8} } \right) \, .
\end{equation}

Let us consider short and long time asymptotics.  If $ \, b \, t \ll 1 $, $ \, C_1
\, $ is determined by a diagram with the minimal number of the interaction lines
(see Fig.\ref{2ColSer}), i.e., we can keep in the power series (\ref{DeltaC})  
the single term with $ \, k = 1 \, $ having $ \, {\cal K}_{\beta} (1) = 1 $, and $
\, {\cal I}_{\beta} (t,1) = \frac{\sqrt{\pi\beta}}{ 2 \, |t| } \, \exp \left( - {
t^2 \over 4 \beta } \right) \, {\rm erf} \left( { t \over \sqrt{4\beta} } \right) $:
\begin{equation}\label{Short}
   C_1( b \, t \ll 1 ) \simeq -N b^{2} \, { \sqrt{\pi\beta} \over 2 }  \, |t| \
                         \exp \left( - { t^2 \over 4 \beta } \right) \, 
                        {\rm erf} \left( { t \over \sqrt{4\beta} } \right) R_N(1) \, ,
\end{equation}
which at $t<\sqrt{4\beta}$ is parametrically smaller than $ \, C_P $: $ 
\, C_1 / C_P |_{ b \, t \ll 1 } \sim b \times ( b \, t ) $. 

One can do the Fourier transform of Eq.(\ref{Short}) and show that for 
large energies $\varepsilon\gg 1$ the correction to the tail of the DOS 
has the same Gaussian exponential dependence as the distribution of 
diagonal matrix elements of $\hat{H}$ unless $R_{N}(1)$ is divergent:
\begin{equation}\label{tail}
  \rho( \varepsilon \gg 1 ) \approx N \sqrt{ \frac{\beta}{2\pi} } \,
       e^{- { \frac{\beta}{2} \varepsilon^{2} } } \,
       \left[ 1 + b^{2} \, \beta \, R_{N}(1) \right].
\end{equation}  
Thus we conclude that there is no slowly decaying Lifshitz tails for almost diagonal
PLBRM with $ \, \alpha > 1/2 \, $ (including the critical PLBRM with $ \, \alpha=1 $) 
even though the multi-point correlation functions may significantly deviate from the 
Poisson distribution.

If $ \, t \gg 1 $, the integral $ \, {\cal I}_{\beta} \, $ can be calculated
approximately by substituting the Dirac $\delta$-function for the exponential in
the integrand of Eq.(\ref{int}):
\begin{equation}\label{int-delta}
   e^{-(tY)^2/\beta} \to \frac{ \sqrt{\beta\pi} }{|t|} \, \delta(Y)
                    \quad \Rightarrow \quad
   {\cal I}_{\beta}(t,k) \simeq \, e^{-\frac{t^2}{4\beta}} \,
         \left( 
             |t|^{-1} \frac{ \sqrt{\beta\pi} }{2^{2k-1} k! \, (k-1)! } + O(1/|t|^{2}) 
         \right) \, ,
\end{equation}
and we obtain the simplified version of the series (\ref{DeltaCsimp}):
\begin{equation}\label{Ser1Dsimp}
   C_1( t \gg 1 ) \simeq - N \, \sqrt{\beta\pi} \, e^{ -\frac{t^2}{4\beta} }
       \sum_{m=1}^{\infty} \! { Z( b \, t, m ) \over |t| } \, 
          e^{ - { Z( b \, t, m ) \over 4 } } \times 
     \left\{
       \begin{array}{l}
         I_0 \Bigl( Z ( b \, t, m ) \Bigr) - I_1 \Bigl( Z ( b \, t, m ) \Bigr) , \ 
                                                                    \beta = 1 \, ; \cr
         1, \ \beta = 2 \, .
       \end{array}
     \right.
\end{equation}
We note that $ \, { Z( b \, t, m ) \over |t| } \sim b \times ( b\, t ) \, $ and,
thus, we can schematically rewrite the asymptotic expression (\ref{Ser1Dsimp}) as
$ \, C_1( t \gg 1 ) \simeq b \times \tilde C_1( b \, t ) $. If the function $ \,
\tilde C_1( b \, t ) \, $ is finite at any value of $ \, ( b \, t ) \, $ the correction 
$ \, C_1 \, $ is again parametrically smaller than $ \, C_P $ (see examples below).

Now we will apply the general formulae (\ref{DeltaC},\ref{DeltaCsimp}) for the
specific models of our interest with the definite correlation function $ \, {\cal
F} ( | i - j | ) $.

{\it The model of Power law banded random matrices}. The model of PLBRM is defined
by Eq.(\ref{VarPL}).  We restrict ourselves to the critical and almost diagonal
PLBRM with $ \, \alpha = 1 \, $ and $ \, b \ll 1 \, $ so that the variance
(\ref{VarPL}) simplifies to the following form [\onlinecite{next}]:
\begin{equation}\label{VarPLsimp}
  \langle | V_{ij}|^{2} \rangle \simeq { 1 \over 2 }
                                  \frac{ b^2  }{  
                        \left( \frac{N}{\pi} \right)^{2} \sin^2 \left( \frac{\pi}{N} |i-j| \right) 
                                               } + O( b^4 ) \, .
\end{equation}
The term of $ \, O( b^4 )  \, $ is not important for the correction $ \, C_1 $. 
We neglect this term below.

We define the correlation function $ \, {\cal F} \, $ in Eq.(\ref{model}) for
critical almost diagonal PLBRM:
\begin{equation}\label{ExmpPL}
    {\cal F}( | i - j | ) = { 1 \over 2 } \, \frac{1}{ 
         \left( \frac{N}{\pi} \right)^{2} \sin^2 \left( \frac{\pi}{N} |i-j| \right)
                                                     } \, , \ i \ne j \, .
\end{equation}

Next we note that the sum in the real space $ \, R_N $, Eq.(\ref{RSS}), is
governed by small distances $ \, m \ll N \, $ and, therefore, its {\it leading
term does not depend on the boundary conditions} for the underlying chain with the
long range hopping:
\begin{equation}\label{PL-simp}
  \sum_{m=1}^{N} {\cal F}^{k} ( m ) = \sum_{m=1}^{N}
                   { 1 \over 2^k } \frac{ 1  }{  
            \left( \frac{N}{\pi} \right)^{2k} \sin^{2k} \left( \frac{\pi}{N} m \right) 
                                          } \simeq \, \sum_{m=1}^{\infty}
             \frac{ 1 }{ \bigl( 2 \, m^{2} \bigr)^k } + O(1/N) \, .
\end{equation}

Substituting Eq.(\ref{PL-simp}) into the series (\ref{DeltaCsimp}) we find the
correction $ \, C_1 $:
\begin{equation} \label{C1pl}
     C_1(t) = - N e^{ -\frac{t^2}{4\beta} } \sum_{m=1}^{\infty} 
        \left( \frac { b \, t }{ m } \right)^2 \, 
        \tilde{{\cal I}}_{\beta} \left[ t, { 1 \over 2 } 
                   \left( \frac { b \, t }{ m } \right)^2 \right] \, .
\end{equation}

Let us consider the limiting cases of small and large time. If $ \, b \, t \ll 1
$, we insert $ \, { {\cal R}_N(1) = \pi^2 / 6 } \, $ into Eq.(\ref{Short}) and
arrive at:
\begin{equation}\label{PLshort}
   C_1( b \, t \ll 1 ) \simeq - N { \pi^{5/2} \sqrt{\beta} \over 12 } b^2 \, |t| \, 
                            \exp \left( - { t^2 \over 4 \beta } \right) \, 
                            {\rm erf} \left( { t \over \sqrt{4\beta} } \right) \, .
\end{equation}
We emphasize that the sum in real space $ \, {\cal R}_N(1) = \sum_m m^{-2} \, $
converges at $ \, m \sim 1 $. Thus, when time $ \, t \, $ is small compared to $
\, b^{-1} \, $ the correction to the Poissonian DOS of the critical PLBRM is
sensitive to the behaviour of the correlation function $ \, {\cal F}(m) \, $ at
short distances. This statement is in fact more general and holds true for any
ensemble of the almost diagonal PLBRM where $ \, \alpha > 1/2 \, $ and $ \, {\cal
R}_N(1) \, $ converges in the thermodynamical limit $ \, N \to \infty $
[\onlinecite{note}].

If $ \, t \gg 1 $, we substitute the correlation function (\ref{ExmpPL}) into the
series (\ref{Ser1Dsimp}) which converges at $ \, m \sim t \, b $. If the product $
\, ( t \, b ) \, $ is large the correction $ \, C_1 \, $ is not sensitive to the
short distances $ \, m \sim 1 \, $ and we can replace the sum over $ \, m \, $
by the integral and find the asymptotics at the long time $ \, t \, b \gg 1 $:
\begin{equation}\label{Ser2Dsimp}
   C_1 ( t \, b \gg 1 ) \simeq - N b \, e^{-\frac{t^2}{4\beta}} 
               \, \times \, 
     \left\{
       \begin{array}{l}
         2,   \ \beta = 1 \, ; \cr
         \pi, \ \beta = 2 \, .
       \end{array}
     \right.
\end{equation}

{\it The Moshe-Neuberger-Shapiro model}. The model of MNS is 
defined by Eq.(\ref{VarMNS}). Let us consider the unitary case with $ \, 
\beta = 2 $. From Eqs.(\ref{VarMNS}) and (\ref{model}) we find the function 
$ \, {\cal F} \, $ for~MNS:
\begin{equation}\label{Fmns}
   {\cal F}( | \phi_i - \phi_j | ) = { 1 \over 2 } \frac{1}{ b^2 + 
                              \left( 
                        \frac{N}{\pi} \sin \left( \frac{\pi}{N} |\phi_i-\phi_j| \right) 
                              \right)^{2}
                                          } \, .
\end{equation}
If $ \, b \ll 1 $, the Gaudin correlation function of the phases $ \, \phi_{i,j}
$, Eq.(\ref{PhR2}), simplifies to the following form:
\begin{equation} \label{PhR2appr}
   {\cal R}_2 ( s/b ) \simeq \frac{(s/b)^2}{ 1 + (s/b)^2} \, .
\end{equation}
The condition $ \, \langle | V_{i,j} |^2 \rangle \ll 1 \, $ holds true in MNS if $
\, | \phi_i - \phi_j | \ge 1 $. However, in contrast to the PLBRM case where the
minimal distance $ \, | \phi_i - \phi_j |=1$, it is violated inside the band $ \, 0
\le | \phi_i - \phi_j | \le b \, $ where the matrix $\hat{H}$ is no longer almost
diagonal. This band is however narrow at $ \, b \ll 1 \, $. Therefore, the
contribution of this band to the DOS averaged over phases $\phi_{i}$ is small
in the parameter $b$.

We apply the strategy of the virial expansion to calculate the average over $ \,
\hat H \, $ at the fixed phases $ \, \phi_i $, see Eq.(\ref{MNSaver}). The phase 
averaging is done at the last step and it reduces the sum in real space to an 
integral over the difference of two phases:
\begin{equation}\label{RSSmns} 
   \langle R_N(k) \rangle_{\phi_i} \simeq 2  \int_{0}^{\infty} \! {\cal F}^k(s) 
                      \, {\cal R}_2(s) \, {\rm d} s \, + O(1/N) \, .
\end{equation}
The case where the correlation function $ \, {\cal F} \, $ depends on the
difference of the integer indices can be restored from
Eq.(\ref{RSSmns}) by substituting a sum of the $\delta$-functions instead of the
two-point correlator: $ \, {\cal R}_2(s) \to \sum_{m=1}^{\infty} \delta( s - m ) $. 
In full analogy with PLBRM we can prove that the leading term of the
function $ \, \langle R_N(k) \rangle_{\phi_i} \, $ in MNS does not depend on the
boundary conditions and transform Eqs.(\ref{Fmns}-\ref{RSSmns}) to a simpler form:
\begin{equation}\label{RSSmnsSimp} 
   \langle R_N(k) \rangle_{\phi_i} \simeq 2  \int_{0}^{\infty} \! \left( 
                 { 1 \over 2 } \frac{1}{ b^2 + s^{2} }
                 \right)^k {\cal R}_2( s ) \, {\rm d} s \, .
\end{equation}
Substituting Eq.(\ref{PhR2appr}) at $ \, b \ll 1 \, $ into formula (\ref{RSSmnsSimp})
we find:
\begin{equation}\label{RSSmnsFin}
   \langle R_N(k) \rangle_{\phi_i} \simeq { 2 \, b \,  \over ( 2 b^2 )^k } 
            \int_{0}^{\infty} \! 
                 \frac{S^2}{ \left( 1 + S^{2} \right)^{k+1} } \ {\rm d} S \, .
\end{equation}
We insert Eq.(\ref{RSSmnsFin}) into series (\ref{DeltaC}) and derive an analog of 
Eq.(\ref{DeltaCsimp}) for MNS:
\begin{equation}\label{C1mns}
  C_1(t) = - N b e^{ -\frac{t^2}{8} } \int_{0}^{\infty} \!\! 
                     { ( t \, S )^2 \over (1 + S^2)^2 } \ \tilde{{\cal I}}_{\beta=2} 
                     \left( t, { t^2 \over 1 + S^2 } \right) \, {\rm d} S \, ,
\end{equation}
which reduces to the following expression:
\begin{equation}\label{MNSansw}
  C_1(t) = - \pi N b \, e^{ -\frac{t^2}{8} }
                       \left( 1 - e^{ -\frac{t^2}{8} } \right) \, .
\end{equation}
The answer (\ref{MNSansw}) coincides with the leading in $ \, b \, $ term of the
Fourier transformed DOS of MNS obtained from the model of noninteracting $ \,
1d \, $ fermions in a parabolic confinement \cite{MNS}.

Let us discuss the results obtained by means of the virial expansion for PLBRM in
GUE (Eq.(\ref{C1pl}) with $ \, \beta = 2 $) and MNS (Eq.(\ref{C1mns})). First we
note that at large time scale, $ \, t \, b \gg 1 $, the summation over the real
space converges at large distances ($ m \sim t b \, $ in PLBRM and $ \, s = S b
\sim t b \, $ in MNS). In this case, we can simplify the integrand in the right
hand side of Eq.(\ref{RSSmnsSimp}) by ignoring $ \, b \, $ in the denominator and
putting $ \, {\cal R}_2 \to 1 \, $. We immediately see that the leading term 
resulting from the sum over $ \, m \, $ in Eq.(\ref{Ser1Dsimp}) is equal to the 
one originating from the integral over $ \, S \, $ in Eq.(\ref{C1mns}). Thus, we
conclude that the first corrections to the Poissonian DOS coincide for PLBRM and
MNS in the long time limit \cite{chi}:
\[
  t \, b \gg 1 \quad \Rightarrow \quad \frac{ C_1( \beta = 2) |_{PLBRM} }
                                            { C_1 |_{MNS} } \simeq 1 \, .
\]

The situation is quite different in the opposite limit of the short time scale $
\, t b \ll 1 \, $ where the asymptotics for $ \, C_1 \, $ is governed by the
single diagram with $ \, k = 1 $. This diagram is highly sensitive to the
behaviour of the function $ \, {\cal F} \, $ at the short distances and,
therefore, yields absolutely different answers for PLBRM and MNS after the
summation in the real space.  At a fixed value of $ \, m \, $ or $ \, s \, $ the
leading diagram is of the order of $ \, \sim b^2 $.  In the case of PLBRM
the sum over $m$ reduces to a numerical prefactor in Eq.(\ref{PLshort}) leaving
the power of $ \, b^2 $ unchanged. The integration over $ \, s \, $ in the
case of MNS is strongly affected by the region $ \, 0 \le s \le b \, $
where $ \, b^2 {\cal F}(s) \, {\cal R}_2(s) |_{s \sim b }\sim 1 $ and the 
off-diagonal elements of $\hat{H}$ are of the order of diagonal ones. This 
region makes the main contribution to the correction to DOS which is small 
only because of the small volume of this region $ \, \Delta s \propto b $.
Thus we find that the integration over small $ \, s \le b \, $ in MNS leads to 
a reduction of the power of $ \, b $ in the short time limit compared to the 
PLBRM case:
\[
  t \, b \ll 1 \quad \Rightarrow \quad \frac{ C_1( \beta = 2) |_{PLBRM} }
                                            { C_1 |_{MNS} } \sim b \, .
\]
This is a clear manifestation of the {\it nonequivalence} of PLBRM 
and MNS models for small $ \, b $, or the nonequivalence of PLBRM at 
small $b$ and the CS model at high temperature.

\begin{acknowledgments}

We are very grateful to Alex Kamenev and Julia Meyer for simulating discussions.

\end{acknowledgments}


\begin{thebibliography}{50}
\bibitem{next} The condition $ \, \alpha = 1 \, $ is not necessary for the application of the
               virial expansion to DOS and it will be removed in the forthcoming paper.
\bibitem{note} If the sum in the real space diverges the virial expansion fails 
               and we cannot consider a finite number of interacting energy 
               levels to derive DOS.
\bibitem{chi} This is similar to the first correction to the Poissonian level 
              compressibility calculated in the paper [\onlinecite{Machin}]. This 
              correction is governed by behaviour of the correlation function $ 
              \, {\cal F} \, $ at the large distances and, therefore, is the same 
              in PLRBM and MNS.
\end{thebibliography}
\end{document}